\documentclass[letterpaper, 10 pt, conference]{ieeeconf}
\usepackage{etex}

\IEEEoverridecommandlockouts                              

\usepackage{times}
\usepackage{latexsym,amsfonts,amsbsy,amssymb}
\usepackage{amsmath}
\usepackage{graphicx}
\usepackage{cite}
\usepackage{algorithm}
\usepackage{algorithmic}
\usepackage{multirow}
\usepackage{mathrsfs}
\usepackage{stmaryrd}
\usepackage{tabularx,booktabs}
\usepackage{xcolor}
\usepackage[curve]{xypic}
\usepackage{subfig}
\usepackage{caption}
\usepackage{color}
\usepackage{tikz}
\usetikzlibrary{automata,positioning,shapes,arrows}

\newtheorem{definition}{Definition}

\newtheorem{problem}{Problem}

\usepackage{float}

\begin{document}
\title{\LARGE \bf Communication-aware Motion Planning for Multi-agent Systems from Signal Temporal Logic Specifications  \thanks{This work was supported by the National Science Foundation (NSF-CNS-1239222 and NSF- EECS-1253488)}}
\author{Zhiyu~Liu,
        Jin~Dai, 
        Bo~Wu
        and Hai~Lin 
\thanks{Z. Liu, J. Dai, B. Wu and H. Lin are with the Department of Electrical Engineering, University of Notre Dame, Notre Dame, IN 46556, USA (e-mail: zliu9@nd.edu; jdai1@nd.edu; bwu3@nd.edu; hlin1@nd.edu).}}
\date{}

\maketitle
\thispagestyle{empty}
\pagestyle{empty}

\begin{abstract}
We propose a mathematical framework for synthesizing motion plans for multi-agent systems that fulfill complex, high-level and formal local specifications in the presence of inter-agent communication. The proposed synthesis framework consists of desired motion specifications in temporal logic (STL) formulas and a local motion controller that ensures the underlying agent not only to accomplish the local specifications but also to avoid collisions with other agents or possible obstacles, while maintaining an optimized communication quality of service (QoS) among the agents. Utilizing a Gaussian fading model for wireless communication channels, the framework synthesizes the desired motion controller by solving a joint optimization problem on motion planning and wireless communication, in which both the STL specifications and the wireless communication conditions are encoded as mixed integer-linear constraints on the variables of the agents' dynamical states and communication channel status. The overall framework is demonstrated by a case study of communication-aware multi-robot motion planning and the effectiveness of the framework is validated by simulation results.
\end{abstract}

\section{Introduction}
Motion planning is a fundamental research problem in robotics and has received a considerable amount of attention in recent years. Traditional planning methods solve the {\it reach-avoid} motion planning problems by taking advantage of various discretized graph search algorithms \cite{choset2005principles}\cite{lavalle2006planning} and randomized sampling-based algorithms \cite{karaman2011sampling}. Feasible motion plans are then constructed for a given model of a robot's dynamics that steer the robot from an initial state to a goal configuration while avoiding obstacles in a complex environment. However, despite the success in dealing with such {\it point-to-point} navigation problems, these approaches lack capability of handling more complex and temporal mission specifications.

Formal languages such as linear temporal logics (LTL) and computation tree logics (CTL) show great potential in specifying and verifying desired complex and logic behavior of systems \cite{baier2008principles}. Incorporating the modern paradigm of hybrid systems with the recent development of formal methods employing temporal logics has allowed us to integrate high-level complex missions with low-level motion controllers \cite{belta2007symbolic}. Based on finite-state abstractions of the dynamics of the robotic system and the environment where it travels, a discrete plan is computed to satisfy the high-level missions by leveraging ideas from formal verification and synthesis \cite{baier2008principles}\cite{kloetzer2008fully}\cite{kress2009temporal} techniques. Such synthesis procedure results in a hybrid control structure with a discrete planner that is responsible for the high-level, discrete plan and a corresponding low-level continuous motion controller. The major limitation of these approaches is their high computational complexity, as both the synthesis and abstraction algorithms scale at least exponentially with the dimension of the discretized configuration space \cite{kloetzer2008fully}. 

Many attempts have been made to apply temporal-logic-based planning techniques to multi-agent cases. Distributed Learning based supervisor synthesis given global temporal specification was studied in \cite{wu2016counterexample}. Filippidis et al. \cite{filippidis2012decentralized} developed a decentralized control scheme for cooperative multi-agent systems from local LTL missions which did not impose any constraints on other agents' behavior. Guo and Dimarogonas \cite{guo2015multi} derived a partially decentralized motion and mission planning solution that decomposed the team into clusters of dependent agents. Applying receding horizon methods \cite{wongpiromsarn2012receding}, Tumova and Dimarogonas \cite{tumova2016multi} further extended the result. On the other hand, multi-agent motion planning from a global specification has also been studied. The vast majority of the existing work in this context focuses on how to properly decompose the global specification into a collection of local tasks, each of which can be fulfilled by individual agents in a synchronized \cite{kloetzer2010automatic} or partially-synchronized \cite{ulusoy2013optimality} manner. Karaman and Frazzoli \cite{karaman2011linear} addressed the mission planning and routing problems for multiple uninhabited aerial vehicles (UAV), in which the given LTL specifications can be systematically converted to a set of constraints suitable to a mixed-integer linear programming (MILP) formulation. Nonetheless, these aforementioned results either assumed perfect inter-agent communication, or reduced the study of communication among the agents to maintenance of topological connectivity of the multi-agent system \cite{guo2015multi}; these assumptions turn out to be oversimplified in practice, since communication quality of service (QoS) does have an impact on multi-agent coordination. Many efforts have been devoted to the communication-aware motion planning problems. Bo et al. \cite{wuADHS2015} developed a combined design framework where the global specification was decomposed into local specifications and artificial potential function based local motion planning was applied considering the communication constraints.  To pursue the co-optimization of motion and communication, Yan and Mostofi \cite{yan2013co}\cite{yan2014go} modeled the communication channel between a robot and a base station as a Gaussian process with fading and shadowing effects; however, the optimization was performed only with respect to the robot's motion velocity, transmission rate and stop time, while the robot was assumed to travel along a pre-defined trajectory.

In addition to the order of control actions in motion planning from formal specifications, we are also concerned with the robustness of the system. Logics with timed features such as metric temporal logic (MTL) \cite{fainekos2009robustness} and signal temporal logic (STL) \cite{maler2004monitoring} have been established to define semantics on real-time signals and to assess the robustness of the systems to parameter or timing variations. In this paper, we adopt STL formulas to characterize the specifications for the multi-agent motion planning problems. STL allows the specification of
properties of dense-time, real-valued signals, and has been applied to the analysis of several types of physical and hybrid systems \cite{maler2004monitoring}. Rather than classical (untimed) temporal logic formulas that justifies the satisfaction of a certain property with binary answers, STL formulas admits a quantitative semantics that provides a real number evaluation that indicates the extent to which the property is satisfied or violated. Recently STL finds applications in controller synthesis for various dynamical systems in either deterministic \cite{raman2014model} or reactive \cite{raman2015reactive} environments.

In this paper, we focus on motion planning problems for multi-agent systems where agents communicate in Gaussian fading channels. The synthesis objective is to construct local controllers to fulfill desired STL local motion tasks of each agent, while optimizing communication QoS between the agents. Our main contribution can be summarized as a unified MILP formalism that solves not only the joint motion-communication co-optimization problem, but synthesizes collision-avoiding motion controllers as well.

The rest of this paper is organized as follows. Section II introduces necessary preliminaries. Section III formally presents the co-optimization problem of communication-aware motion planning from STL specifications. The MILP encoding of communication-aware motion planning is studied in Section IV for the purpose of motion controller synthesis. Simulation examples are presented in Section V. Section VI concludes the paper.

\section{Preliminaries}
\subsection{Agent Dynamics}
We consider $P$  agents with unique identities $\mathcal{P}=\{1,2, \ldots, P\}$ that perform their motions in a shared 2-D environment. For each $i\in \mathcal{P}$, the motion of agent $i$ is captured by the linear dynamics of the following form
\begin{equation}
\dot x_i(t)=Ax_i(t)+Bu_i(t),
\label{dynamic}
\end{equation}
where $x_i\in\mathbb{R}^4$ is the state of agent $i$ with $x_i=[p_i^T\quad v_i^T]^T$, where $p_i, v_i\in\mathbb{R}^2$ denote the position and velocity of the agent, respectively; $u_i=[u_{i,1}\quad u_{i,2}]^T\in \mathcal{U}\subseteq \mathbb{R}^2$ is the local admissible control inputs, and $x_i(0)=x_{i,0}\in \mathbb{R}^4$ is the initial state. $A$ and $B$ are matrices with proper dimensions, and $(A,B)$ is a controllable pair. The environment $\mathcal{X}$ shared by the agents is given by a large convex polygonal subset of the $2$-D Euclidean space $\mathbb{R}^2$. Let $\mathcal{X}_{obs}\subseteq \mathcal{X}$ denote the regions in the environment that are occupied by (polygon) obstacles. $\mathcal{X}_{free}=\mathcal{X}\setminus\mathcal{X}_{obs}$ denotes the obstacle-free working space for the multi-agent system.

To pursue the communication-aware motion planning in an online manner, we follow up the idea from \cite{raman2015reactive} and assume that the robot dynamics (1) admits a discrete-time approximation of the following form, given an appropriate sampling time $\Delta t>0$:
\begin{equation}
x_i(t_{k+1})=A_dx_i(t_{k})+B_du_i(t_{k}),
\end{equation}
where $k\in\mathbb{N}$ is the sampling index and $\Delta t$ is selected such that $(A_d, B_d)$ is controllable. Note that the sampling is uniformly performed, i.e., for each $k>0$, $t_{k+1}-t_k=\Delta t$. For simplicity, we slightly abuse the notations and use $[a,b]$ as an abbreviation for the set $\{a,a+1,\ldots,b\}$.

Given $x_{i,0}\in\mathbb{R}^2$ and $\bf{u}_i\in \mathcal{U}^\omega$, $i\in\mathcal{P}$, a (state) {\it run} ${\bf x}_i= x_{i,0}x_{i,1}x_{i,2}\ldots$ generated by agent $i$'s dynamics (2) with control input $\bf {u_i}$ is an infinite sequence obtained from agent $i$'s state trajectory, where $x_{i,k}=x_i(t_k)\in\mathbb{R}^4$ is the state of the system at time index $t$, and for each $k\in\mathbb{N}$, there exists a control input $u_{i,k}=u_i(t_k)\in \mathcal{U}$ such that $x_i(t_{k+1})=A_dx_i(t_{k})+B_du_i(t_{k})$. Given a local initial state $x_{i,0}$ and a sequence of local control inputs ${\bf u}^N_i = u_{i,0}u_{i,1}u_{i,2}\ldots u_{i,N-1}$, the resulting horizon-$N$ run of agent $i$, ${\bf x}_i(x_{i,0},{\bf u}^N_i)=x_{i,0}x_{i,1}x_{i,2}\ldots x_{i,N}$ is unique.

\subsection{Signal Temporal Logic}
We consider STL formulas that are defined recursively as follows.

\begin{definition}[STL Syntax]\rm
STL formulas are defined recursively as:
$$
\varphi::={\rm True}|\pi^\mu|\neg\pi^{\mu}|\varphi\land\psi|\varphi\lor\psi|\Box_{[a,b]} \psi | \varphi\sqcup_{[a,b]} \psi
$$
\end{definition}
where $\pi^\mu$ is an atomic predicate $\mathbb{R}^n\to\{0,1\}$ whose truth value is determined by the sign of a function $\mu:\mathbb{R}^n\to\mathbb{R}$, i.e., $\pi^\mu$ is true if and only if $\mu({\bf x})>0$; and $\psi$ is an STL formula. The ``eventually" operator $\Diamond$ can also be defined here by setting $\Diamond_{[a,b]} \varphi={\rm True}\sqcup_{[a,b]} \varphi$.

The semantics of STL with respect to a discrete-time signal $\bf x$ are introduced as follows, where $({\bf x},t_k)\models \varphi$ denotes for which signal values and at what time index the formula $\varphi$ holds true.
\begin{definition}[STL Semantics]\rm
The validity of an STL formula $\varphi$ with respect to signal $\bf x$ at time $t_k$ is defined inductively as follows:
\begin{enumerate}
\item $({\bf x},t_k)\models \mu$, if and only if $\mu(x_k)>0$;
\item $({\bf x},t_k)\models \neg\mu$, if and only if $\neg(({\bf x},t_k)\models \mu)$;
\item $({\bf x},t_k)\models \varphi\land\psi$, if and only if $({\bf x},t_k)\models \varphi$ and $({\bf x},t_k)\models \psi$;
\item $({\bf x},t_k)\models \varphi\lor\psi$, if and only if $({\bf x},t_k)\models \varphi$ or $({\bf x},t_k)\models \psi$;
\item $({\bf x},t_k)\models \Box_{[a,b]}\varphi$, if and only if $\forall t_{k'}\in[t_k+a,t_k+b]$, $({\bf x},t_{k'})\models \varphi$;
\item $({\bf x},t_k)\models \varphi\sqcup_{[a,b]}\psi$, if and only if $\exists t_{k'}\in[t_k+a,t_k+b]$ such that $({\bf x},t_{k'})\models \psi$ and $\forall t_{k''}\in[t_k,t_{k'}]$, $({\bf x},t_{k''})\models \varphi$.
\end{enumerate}
\end{definition}

A signal ${\bf x}= x_0x_1x_2\ldots$ satisfies $\varphi$, denoted by $\bf x\models\varphi$, if $({\bf x},t_0)\models\varphi$.

Intuitively, $\bf x\models \Box_{[a,b]}\varphi$ if $\varphi$ holds at every time step between $a$ and $b$, ${\bf x}\models \varphi\sqcup_{[a,b]}\psi$ if $\varphi$ holds at every time step before $\psi$ holds, and $\psi$ holds at some time step between $a$ and $b$, and ${\bf x}\models \Diamond_{[a,b]}\varphi$ if $\varphi$ holds at some time step between $a$ and $b$.

An STL formula $\varphi$ is {\it bounded-time} if it contains no unbounded operators. The bound of $\varphi$ can be interpreted as the horizon of future predicted signals $\bf x$ that is needed to calculate the satisfaction of $\varphi$.

\subsection{Communication Channel}
Inter-agent communication is considered at this point. In particular, we consider the average bit error rate (BER) among the agents. Since BER directly depends on the received signal strength, we build our channel model by estimating the received signal strength index (RSSI) \cite{fink2013robust}. Given $N$ training samples of RSSI for a source and receiver pair: $z_{i}$ and $y_{i}=(p_{i,s},p_{i,r})$ where $i\in [1, K]$, we denote $Y_T={y_{1},\cdots,y_{K}}$ for positions and $Z_T={z_{1},\cdots,z_{K}}$ for RSSI measurements, where $T$ stands for training.

To estimate the communication channel considering fading and shadowing effects, we employ the spatial Gaussian process ($\mathcal{GP}$) model \cite{fink2013robust}. The distribution of the RSSI for a sender-receiver pair can be expressed as a Gaussian Process as follows.

\begin{equation}
\begin{aligned}
RSSI(y) \sim \mathcal{GP}(&m(y)+C_{y,T}[C_T+\sigma_F^2I]^{-1}(z_T-m(Y_T)), \\
&C_y-C_{y,T}[C_T+\sigma_F^2I]^{-1}C_{T,y}+\sigma_F^2I),
\end{aligned}
\end{equation}
where $\sigma_F^2$ models fading effect. Combining the path loss model in \cite{yan2014go}, we choose $$m(y)=L_0-10n_llog_{10}(\left \| p_s-p_r \right \|),$$
where $L_0$ is the received power at 1 m from the source and $n_l$ is a path-loss exponent.
Let
$$C(y,y')=\sigma_{k}^2 e^{\dfrac{-d(y,y')^2}{2l^2}}$$
describe the shadowing effect, where $\sigma_{k}^2$ and $l$ are related to shadowing effects. $C_{y,T}$ is the covariance vector between predicting points and training samples and $C_T$ is the covariance matrix of training samples. The function $d(y,y')$ denotes the distance between $y$ and $y'$, and is selected to be:

\begin{equation}
\resizebox{0.8\hsize}{!} {$d(y,y')=min\left \{ \left \| \binom{p_s}{p_r}-\binom{p_s'}{p_r'} \right \|,\left \| \binom{p_s}{p_r}-\binom{p_r'}{p_s'} \right \|\right \}$}
\end{equation}

  Given a few training samples we use maximum likelihood estimator to find the best estimation for these hyper parameters in terms of probability. The maximum likelihood estimator of RSSI is:
\begin{equation}\label{eqn:RSSI}
\begin{aligned}
&\overline{RSSI}(y)_{dB} \\
&=m(y)+C_{y,T}[C_T+\sigma_F^2I]^{-1}(z_T-m(Y_T))\\
&=L_0-10n_llog_{10}(||p_s-p_r||)\\
&+\resizebox{0.8\hsize}{!} {$\begin{bmatrix}C(y,y_1)\\ \vdots\\ C(y,y_K) \end{bmatrix}^T\begin{bmatrix}
	C(y_1,y_1)+\sigma_F^2 & \cdots &C(y_K,y_1) \\
	C(y_1,y_2) & \cdots &\vdots \\
	\vdots & \cdots & C(y_K,y_K)+\sigma_F^2
	\end{bmatrix}^{-1}\begin{bmatrix}
	z_1-m(y_1)\\
	\vdots\\
	z_K-m(y_K)
	\end{bmatrix}$}
\end{aligned}
\end{equation}


\section{Problem Formulation}
\subsection{STL Motion Planning Specifications}
We now proceed to the communication-aware motion planning problems for multi-agent systems. Let us consider a team of $P$ agents conducting motion behavior in the shared environment $\mathcal{X}$, each of which is governed by the discretized dynamics (2). We assign a goal region $\mathcal{X}_{i,goal}$ for agent $i$, $i\in\mathcal{P}$ that is characterized by a {\it polytope} \cite{kloetzer2008fully} in $\mathcal{X}_{free}$, i.e., there exist $M\ge3$ and $a_{i,j}\in \mathbb{R}^2$, $b_{i,j}\in\mathbb{R}$, $j=1,2,\ldots, M_i$ such that
\begin{equation}
\mathcal{X}_{i,goal}=\{p\in\mathbb{R}^2|a_{i,j}^Tp+b_{i,j}\le 0, j=1,2,\ldots, M_i\}.
\end{equation}

In other words,
\begin{equation}
\mathcal{X}_{i,goal}=\{x\in\mathcal{X}|a_{i,j}^T[I_2\quad O_2]x+b_{i,j}\le 0, j=1,2,\ldots, M_i\}.
\end{equation}
where $I_2, O_2\in\mathbb{R}^{2\times 2}$ denote the 2-dimensional identity and zero matrices, respectively.

Without loss of generality, we also assume that the region $\mathcal{X}_{obs}$ is  a polygonal subset of $\mathcal{X}$, i.e., there exist an integer $M_{obs}\ge 3$ and $a_{obs,j}\in \mathbb{R}^2$, $b_{obs,j}\in\mathbb{R}$, $j=1,2,\ldots, M_{obs}$ such that
\begin{equation}
\mathcal{X}_{obs}=\{p\in\mathbb{R}^2|a_{obs,j}^Tp+b_{obs,j}\le 0, j=1,2,\ldots, M_{obs}\}.
\end{equation}

We assume that all agents share a synchronized clock. The terminal time of multi-agent motion is upper-bounded by $t_f=T_f\Delta t$ with some $T_f\in\mathbb{N}$, and the planning horizon is then given by $[0,T_f]$. Accomplishment of individually-assigned specifications is of practical importance, for instance search and rescue missions or coverage tasks are often specified to mobile robots individually. In this paper, local motion planning tasks for agent $i$ are summarized as the following STL formula: for $i\in\mathcal{P}$, require:

\begin{equation}
\varphi_i=\varphi_{i,p}\land\varphi_{i,s},
\end{equation}
where
\begin{enumerate}
\item the motion performance property
\begin{equation}
\varphi_{i,p}=\Diamond_{[0,T_f]} \bigwedge_{j=1}^{M_i}\left(a_{i,j}^T[I_2\quad O_2]x_i+b_{i,j}\le 0\right)
\end{equation}
requires that agent $i$ enter the goal region within $T_f$ time steps;
\item the safety property
\begin{equation}
\begin{aligned}
&\varphi_{i,s}=\varphi_{i,s,obs}\land\varphi_{i,s,col}\\
&=\left[\Box_{[0,T_f]}\bigwedge_{j=1}^{M_{obs}}\left(a_{obs,j}^T[I_2\quad O_2]x_i+b_{obs,j}> 0\right)\right]\bigwedge\\
&\left[\Box_{[0,T_f]} \bigwedge_{j\in\mathcal{P}, j\ne i}(|p_{i,1}-p_{j,1}|\ge d_1)\land(|p_{i,2}-p_{j,2}|\ge d_2)\right]
\end{aligned}
\end{equation}
ensures that agent $i$ shall never encounter obstacle regions nor collide with other agents. Here $d_1$ and $d_2$ are pre-defined safety distances between two agents in the two dimensions.
\end{enumerate}
\subsection{Communication and Motion Co-optimization Problem}
We aim to steer the agents to not only satisfy the local motion specifications, but to optimize the energy consumptions and inter-agent communication QoS as well. Towards this end, we adopt cost functions in the following linear quadratic form to represent the total energy consumption of the underlying multi-agent system.
\begin{equation}\label{eqn:J1}
J_{1}=
\sum_{k=0}^{T_f}\sum_{i=1}^{P}({q}^T|{x_{i,t_k}|+{r}^T|u_{i,t_k}|})
\end{equation}

where ${q}$, ${r}$ are non-negative weighting column vectors and $|.|$ denotes the element-wise absolute value such that the cost can be encoded by MILP. On the other hand, we consider the cost function that accounts for the communication QoS. As we can see in the previous section, (\ref{eqn:RSSI}) is highly non-linear with respect to the position pair $y$. To make the problem solvable in MILP, we need to linearize the communication cost and constraints. To this end, we divide the working space into $n$ partitions and denote the matrix $G=[\frac{1}{RSSI}_{ij}]$ where $RSSI_{ij}$ is the expected RSSI from (\ref{eqn:RSSI}) between two centers of the partitions $i$ and $j$ and does not equal to $0$dB. We assume $G_{i,j}$ sufficiently approximates the RSSI from any point in partition $i$ to any point in partition $j$. Furthermore, we define the binary matrix $O_t$, $t\in[0,T_f]$ to capture the occupancy of the partitions where $O_{i,j,t}$ is zero if and only if the sender agent is in partition $i$ and the receiver agent is in partition $j$ at time $t$. The dimensions of $G$ and $O_t$ are $n\times n$. Then $J_2$ defined as below characterizes the cost of communication between the agents as they move towards their goals.

\begin{equation}\label{eqn:J2}
J_{2}= \sum_{k=0}^{T_f}\sum_{i=1}^{n}\sum_{j=1}^{n}G_{i,j}(1-O_{i,j,t_k})
\end{equation}

Based on the aforementioned preliminaries and cost functions, we now formally state the communication and motion co-optimization problem from STL specifications as follows.

\begin{problem}[Communication and Motion Co-optimization]
Given a multi-agent system that consists of $P$ interacting agents, each of which is governed by a discrete-time dynamics (2) and is initially associated with an initial state $x_{i,0}$, a planning horizon $T_f$, and a local STL specification formula $\varphi_i$ in (9), compute local control inputs $u_i(t_k)$, $i\in\mathcal{P}$, $k\in[0,T_f]$ such that the following convex hull of cost functions $J_1$ and $J_2$ is optimized ($0<\alpha<1$):
\begin{equation}
\underset{{\bf u}_i^{T_f}, i\in\mathcal{P}}{\min}\quad J({\bf x}_i(x_{i,0},{\bf u}_i^{T_f}))=\alpha J_1+(1-\alpha)J_2
\end{equation}
\begin{alignat*}{3}
\mbox{s.t.}\quad &\forall i\in\mathcal{P}, \\
&x_i(t_{k+1})=A_dx_i(t_k)+B_du_i(t_k),\\
&{\bf x}_i(x_{i,0},{\bf u}_i^{T_f}) \models \varphi_i,\\
&u_i\in\mathcal{U}=[-u_{max}, u_{max}]\times[-u_{max}, u_{max}], \\
&||v_i||<v_{max},\\
& \omega_i=\frac{||u_i||}{m_i||v_i||}\leq\frac{u_{max}}{m_iv_{max}}
\end{alignat*}
where $u_{max}$ and $v_{max}$ are constants that bound $u_i$ and $v_i$, $\omega_i$ is the turning rate, $m_i$ denotes the mass of agent $i$.
\end{problem}
\subsection{Overview of the MILP Paradigm}
We propose a two-layer planning and synthesis framework to solve Problem 1 for the underlying multi-agent system.

\begin{itemize}
\item The top layer consists of two MILP encoding processes. On one hand, we introduce a Boolean variable $z_{t_k}^{\varphi_i}$ for agent $i$ to justify whether or not $\varphi_i$ is satisfied at time step $t_k$, which is explained in the following section. On the other hand, to achieve communication-aware motion planning, the inter-agent communication model that is illustrated in (3)-(5) is also characterized as mixed integer-logical constraints that are related to the optimization of the cost function (14).
\item The bottom layer is an MILP solver that solves Problem 1 by converting it to an MILP problem with mixed integer-logical constraints from the two perspectives in the top layer. The MILP solver works out feasible and performance-optimized motion paths for each agent that consist of a series of waypoints and local control inputs.
\end{itemize}

\section{MILP Encoding of Communication-aware Motion Planning}

\subsection{MILP Encoding of Agent Dynamics}
For sake of simplicity, we replace $t_k$ with $t$ and denote $u_{it}$ and $x_{it}$ as the control input and state for agent $i$ at time step $t$, respectively. For the motion and control cost $J_1$ defined in (\ref{eqn:J1}), we transform this convex, piece-wise cost into a linear form by introducing slack vectors $\alpha_{it}$ and $\beta_{it}$ and the additional constraints \cite{athans2013optimal}.
\begin{equation}
J_{1}=\sum_{t=1}^{T_f}\sum_{i=1}^{P}({q}^T{\alpha_{it}+{r}^T\beta_{it}})
\end{equation}

\begin{equation}
\begin{aligned}
&\text{s.t.} & \forall t\in [1,&T_f], \forall i\in [1, P], \forall j\in [1,4], \forall k\in [1, 2] \\
& \text{}
& x_{itj}&\leq \alpha_{itj}, -x_{itj}\leq \alpha_{itj} \\
& \text{and}
& u_{itk}&\leq \beta_{itk}, -u_{itk}\leq \beta_{itk}\\
& \text{and}
& x_i(t+1)&=A_dx_i(t)+B_du_i(t)
\end{aligned}
\end{equation}

The given velocity constraints are nonlinear, we use an arbitrary number $H$
of linear constraints to approximate it \cite{richards2002aircraft}. The 2-D velocities are bounded by a regular H-sided polygon.
\begin{equation}
\begin{split}
\forall h\in [1,H], i\in[1,P],t\in[1,T_f]\\
v_{it1}sin(\frac{2\pi h}{H})+v_{it2}cos(\frac{2\pi h}{H})\leq v_{max}
\end{split}
\end{equation}

\subsection{Boolean Encoding of STL Constraints}
 Using the method in \cite{raman2014model}, the MILP encoding of local motion planning specification $\varphi_i$ for agent $i$, $i\in\mathcal{P}$, relies on three Boolean variables, namely $z_{t}^{\varphi_{i,p}}$, $z_{t}^{\varphi_{i,s,obs}}$ and $z_{t}^{\varphi_{i,s,col}}$, that correspond to the satisfaction of $\varphi_{i,p}$, $\varphi_{i,s,obs}$ and $\varphi_{i,s,col}$, respectively. We introduce another Boolean variable $z_{t}^{\varphi_i}$ whose truth value determines the satisfaction of $\varphi_i$ at time $t$, by combining the encoded constraints for $z_{t}^{\varphi_{i,p}}$ and $z_{t}^{\varphi_{i,s,obs}}$ and $z_{t}^{\varphi_{i,s,col}}$.

\begin{equation}
z_{t}^{\varphi_i}=z_{t}^{\varphi_{i,p}}\land z_{t}^{\varphi_{i,s,obs}} \land z_{t}^{\varphi_{i,s,col}}
\end{equation}
with
\begin{equation}
\begin{split}
\forall i\in\mathcal{P}: \\
&z_{t}^{\varphi_i}\le z_{t}^{\varphi_{i,p}}, z_{t}^{\varphi_i}\le z_{t}^{\varphi_{i,s,obs}},z_{t}^{\varphi_i}\le z_{t}^{\varphi_{i,s,col}}\\
&z_{t}^{\varphi_i}\le z_{t}^{\varphi_{i,p}}+z_{t}^{\varphi_{i,s,obs}}+z_{t}^{\varphi_{i,s,col}}-2
\end{split}
\end{equation}
where $z_{t}^{\varphi_{i,p}}$, $z_{t}^{\varphi_{i,s,obs}}$ and $z_{t}^{\varphi_{i,s,col}}$ are one if and only if their corresponding specifications are satisfied.

\subsection{MILP Encoding of Communication Constraints}


%

For the communication cost, assuming that we divide the work space into $n=N^2$ small grids with $N$ rows and $N$ columns and each partition has size $d \times d$, we use big-M formulation \cite{griva2009linear} to describe the linear constraints as the following (assuming two agents).

\begin{equation}
\begin{aligned}
\forall &i,j\in[1,n],p,q \in[1,P],t\in[1,T_f]\\
x_{pt1} &\geq x_{min}+(a_{pt}-1)d, x_{pt1} \leq x_{min}+a_{pt} d\\
x_{pt2} &\geq y_{min}+(b_{pt}-1)d, x_{pt2} \leq y_{min}+b_{pt} d \\
r_{pt} &= (a_{pt}-1)  N + b_{pt}, r_{qt} = (a_{qt}-1)  N + b_{qt} \\
(i&-r_{pt})-MO_{ijt} \leq 0, -(i-r_{pt})-MO_{ijt} \leq 0 \\
(j&-r_{qt})-MO_{ijt} \leq 0, -(j-r_{qt})-MO_{ijt} \leq 0 \\
\sum_{i,j} &O_{ijt}=n^2-1
\end{aligned}
\end{equation}
where $M$ is a large number; $x_{min}$ and $y_{min}$ are the minimum coordinates of the working space; $a_{pt}$, $b_{pt}$ are integers from 1 to $N$ representing the rows and columns position of agent $p$ in the working space; $r_{pt}$ and $r_{qt}$ are integers from 1 to $n$ representing the partitions where agent $p$ and $q$ locate.

\section{Simulation Results}
Based on the MILP formulation of both the STL specifications and the communication-awareness, we aim to test our co-optimization strategy. For such a purpose, we ran simulations in MATLAB and employed AMPL/Gurobi to solve the optimization problem. A Mathematical Programming Language (AMPL) is an algebraic modeling language to describe and solve high-complexity problems for large-scale mathematical computing \cite{WinNT}. Gurobi Solver, a commercial optimization solver for MILP, finds optimal solutions to the problem formulated by AMPL.

\subsection{Motion Planning using MILP}
\begin{figure}[!t]
	\centering
	\includegraphics[width=1.1\linewidth]{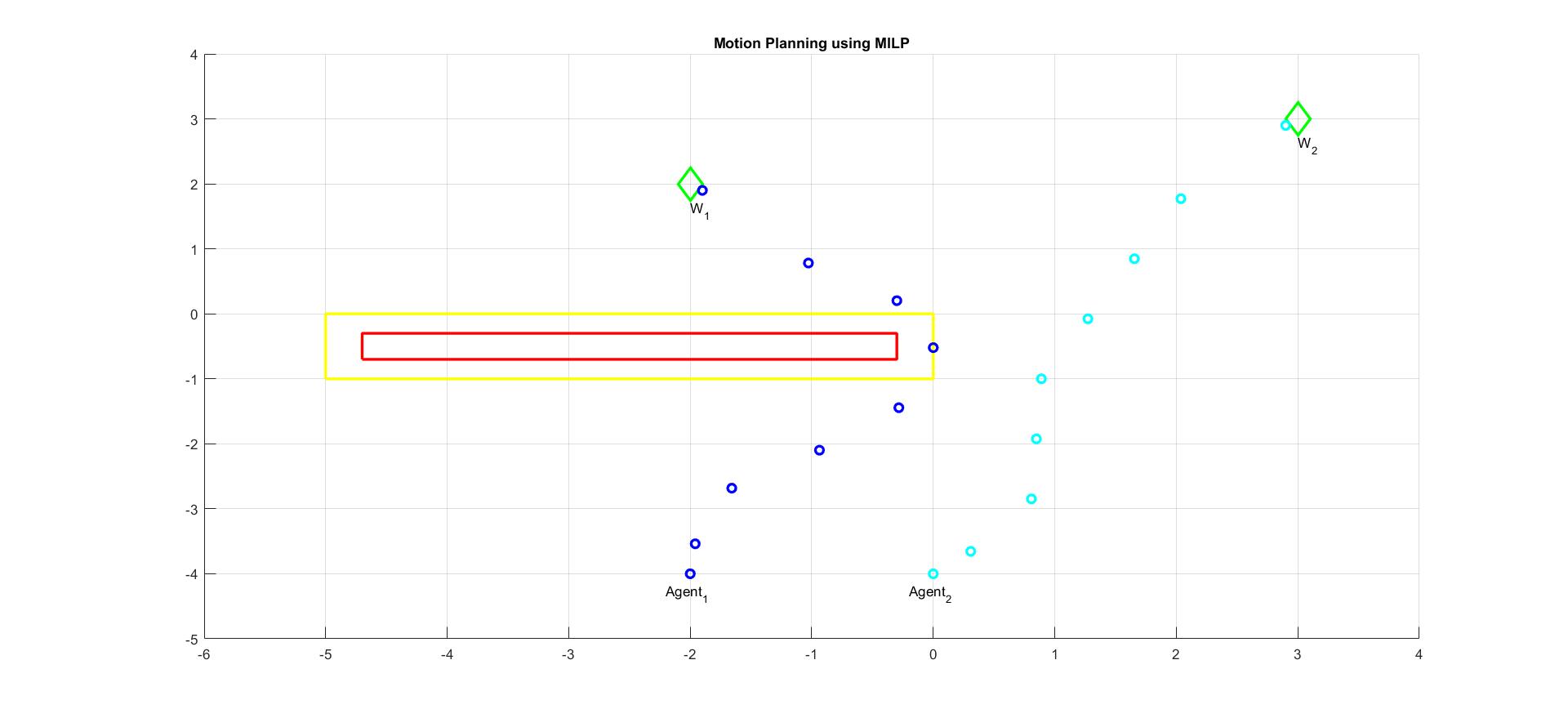}
	\caption{Motion Planning using MILP}
	\label{fig_1}
\end{figure}

To run the motion planning using MILP, we set $P=2$, $H=8$, $M=99999$ and $T_f=8$. The matrices $A$ and $B$ in the discretized dynamics (2) of each agent are given by
\begin{equation}
A_d=
\begin{bmatrix}
1 &0 &1 &0 \\
0 &1 &0 &1 \\
0 &0 &1 &0 \\
0 &0 &0 &1
\end{bmatrix},
B_d=
\begin{bmatrix}
0.5 &0\\
0 &0.5\\
1 &0\\
0 &1
\end{bmatrix}.
\end{equation}
Considering the actual size of the agents, we set a buffer box (yellow rectangle in Fig. \ref{fig_1}) for the obstacle which is larger than its actual size. The red rectangle is the real obstacle by setting $a_{obs,1}=-4.7$, $a_{obs,2}=-0.7$, $b_{obs,1}=-0.3$ and $b_{obs,2}=-0.3$. The output of simulation is the position, velocity and control input of each agent at all steps. As we can see from the Fig. \ref{fig_1}, given initial and target position of two agents and one static obstacle, the agent is able to arrive at its destination without obstacle collision using MILP under AMPL/Gurobi.

\subsection{Communication-aware Motion Planning}
\begin{figure}[!t]
	\centering
	\includegraphics[width=1.1\linewidth]{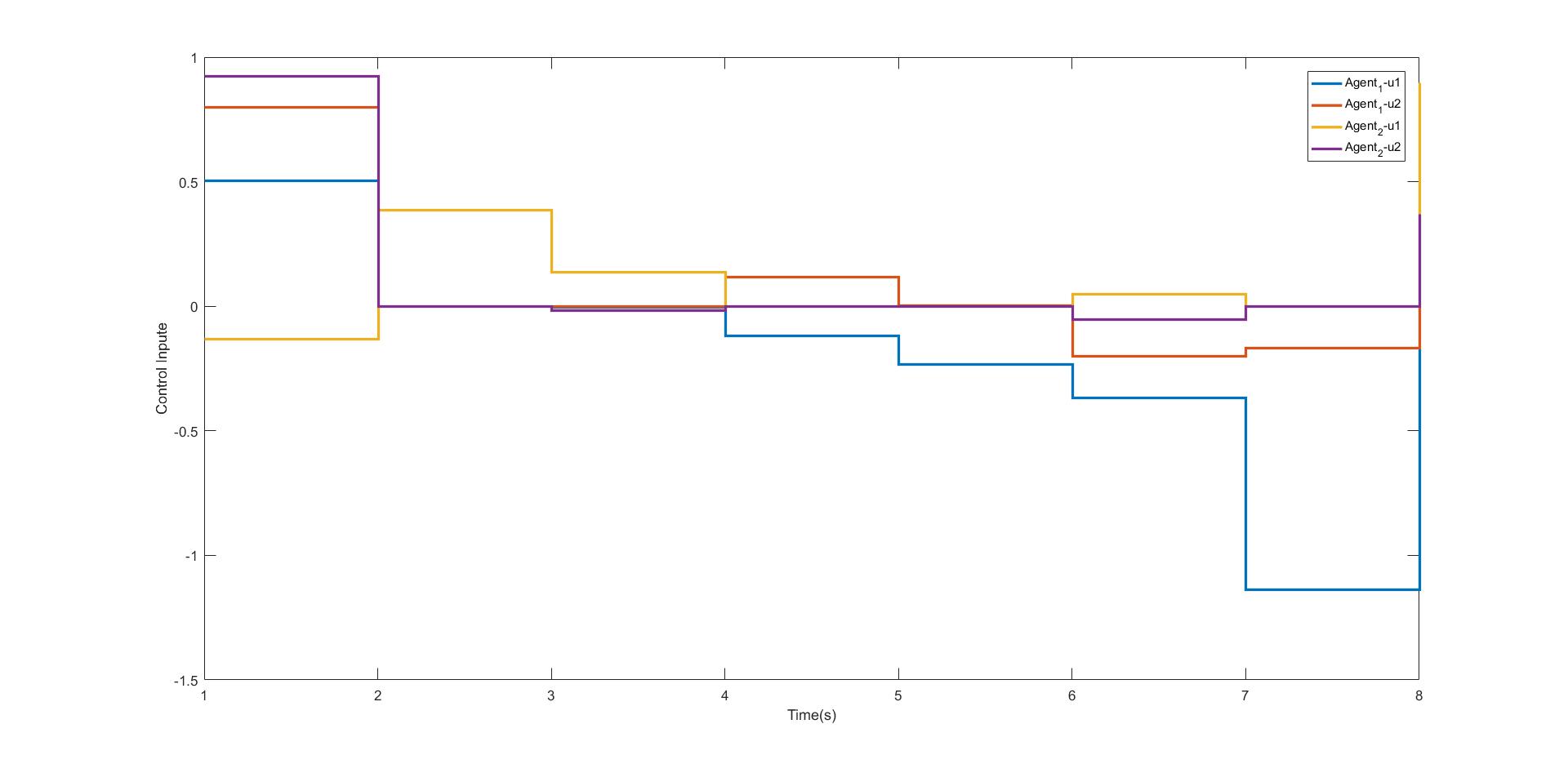}
	\caption{Control Input of Communication-aware Motion Planning}
	\label{controlinput}
\end{figure}

\begin{figure}[!t]
	\centering
	\includegraphics[width=1.1\linewidth]{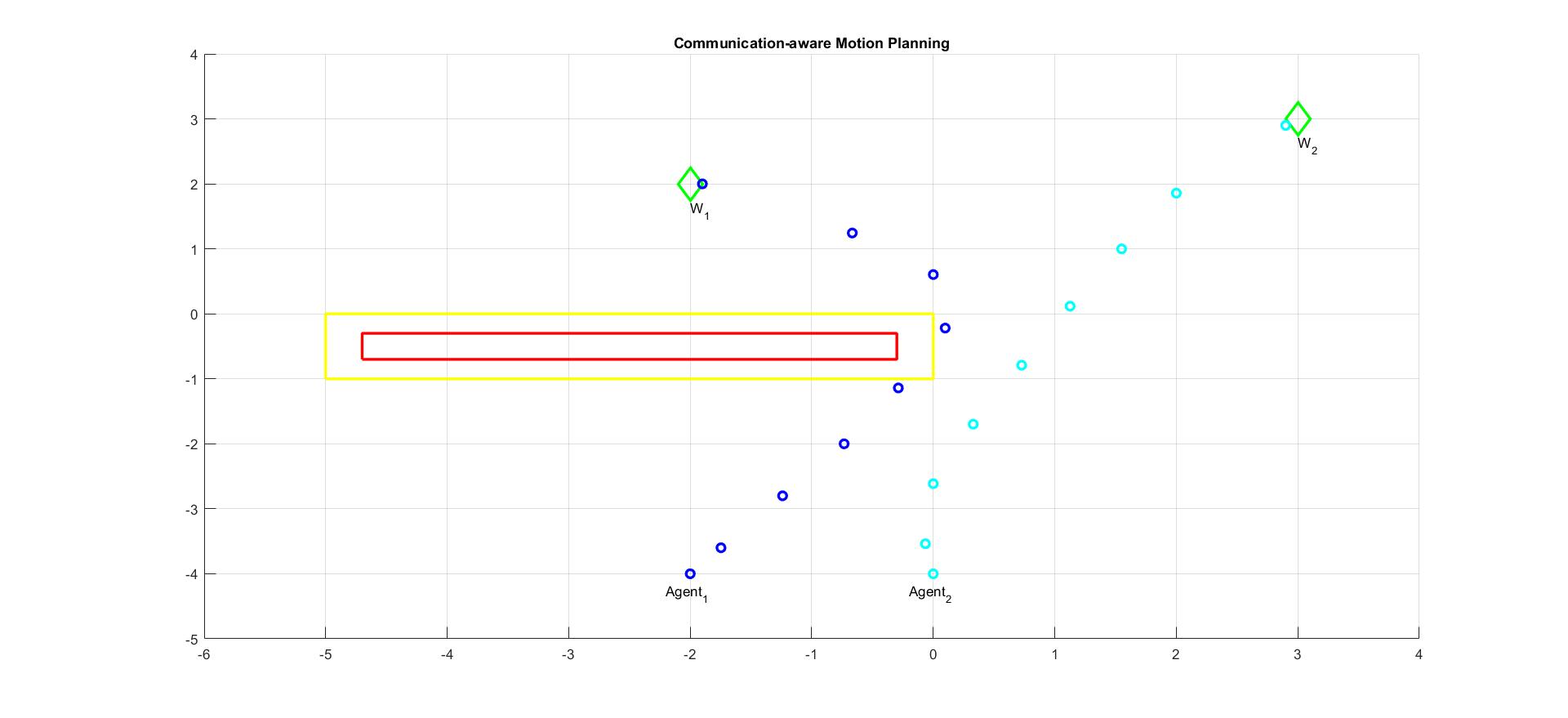}
	\caption{Communication-aware Motion Planning}
	\label{fig_2}
\end{figure}

We added communication cost into consideration within the same working environment and set the hyper parameters for the communication model as: $L_0=-12.89dB$, $n_l=3$, $\sigma_F^2=10$, $\sigma_{k,1}^2=10$, $l_1=2$ \cite{gonzalez2012comprehensive}. We chose the resolution of each small grid as 1 meter, where the size of the working space was $10m \times 10m$ like the case above and chose coefficient $\alpha$ as 0.1. After AMPL formulated the optimization problem, we had 73327 variables and 291808 constraints (all linear). We ran the simulation on a PC with Intel core i7-4710MQ 2.50 GHz processor and 8GB RAM. It took 783 seconds to solve the problem. As in the first example, the output of the simulation is the position, velocity and control input of each agent at all steps. The control input signals are shown in Fig. \ref{controlinput}. The motion planning results from the joint optimization is shown in Fig. \ref{fig_2}. Compared to the case in which communication evaluation was ignored, the two agents are closer. The total cost is 22.296 where the communication part is 14.216. We also calculate the communication cost for the first case, which is 23.840, 67.7\% larger than the second case. The communication quality has been improved by our joint optimization strategy.

\section{Conclusion}
The communication-aware motion planning problem for multi-agent systems is considered in this paper. By specifying local motion tasks as signal temporal logic formulas and modeling inter-agent communication as Gaussian channels, we propose a co-optimization framework that optimizes the total energy consumption of the agents and communication QoS among the agents simultaneously, while guaranteeing the accomplishment of each agent's motion specifications. A mixed integer-logical programming formalism is deployed to explore both satisfaction of STL specification and communication-motion co-optimization. Effectiveness of the proposed framework is validated by a 2-agent motion planning simulation.

\bibliographystyle{IEEEtran}
\bibliography{comm}
\end{document}